# X-ray Photoemission Study of MgB$_2$

R. P. Vasquez*

Center for Space Microelectronics Technology, Jet Propulsion Laboratory, California Institute of Technology, Pasadena, California  91109-8099

C. U. Jung, Min-Seok Park, Heon-Jung Kim, J. Y. Kim, and Sung-Ik Lee

National Creative Research Initiative Center for Superconductivity and Department of Physics, Pohang University of Science and Technology, Pohang 790-784,  Republic of Korea

## Abstract

A c-axis oriented thin film and a high density sintered pellet of MgB$_2$ have been studied by x-ray photoemission spectroscopy, and compared to measurements from MgO and MgF$_2$ single crystals.  The as-grown surface has a layer which is Mg-rich and oxidized, which is effectively removed by a nonaqueous etchant.  The subsurface region of the pellet is Mg-deficient.  This nonideal near-surface region may explain varied scanning tunneling spectroscopy results.  The MgB$_2$ core level and Auger signals are similar to measurements from metallic Mg and transition metal diborides, and the measured valence band is consistent with the calculated density of states.

*email: richard.p.vasquez@jpl.nasa.gov

## 1. Introduction

The recent discovery[1] of superconductivity at 39 K in $MgB_2$ has generated a great deal of interest. This is the highest reported superconducting transition temperature ($T_c$) outside of the cuprate high temperature superconductors and certain doped fullerenes. Like the cuprates, $MgB_2$ appears to be hole-doped, as is evident from theoretical considerations[2-5] and Hall effect measurements,[6] and has a layered crystal structure, with planar $sp^2$ bonding yielding a structure very similar to graphite. However, unlike the cuprates $MgB_2$ appears to be a conventional BCS superconductor, based on early evidence from $^{11}B$ NMR measurements,[7] the B isotope effect,[8] and scanning tunneling spectroscopy (STS) measurements.[9-13] These indications of conventional superconductivity are preliminary, and the STS measurements in particular appear to have problems of reproducibility, most likely due to variations in surface quality. Thus, it is important to characterize the initial surface, to understand mechanisms of surface degradation, and to develop methods of obtaining clean surfaces capable of yielding measurements intrinsic to the superconducting phase.

In this work, $MgB_2$ surfaces are characterized with x-ray photoemission spectroscopy (XPS). The core levels, Mg KLL Auger transitions, and valence bands measured from a c-axis oriented thin film and a sintered polycrystalline pellet of $MgB_2$ are compared to XPS measurements from MgO (100) and $MgF_2$ (100) crystal surfaces. Nonaqueous chemical etching is shown to effectively remove oxidized Mg from the $MgB_2$ surface, and XPS signals intrinsic to the superconducting phase are identified. This work will show that the Mg core level and Auger signals of $MgB_2$ are similar to measurements from metallic Mg, the B 1s signal is similar to measurements from transition metal diborides, and the measured valence band is consistent with the calculated density of states.

## 2. Experimental

A c-axis oriented thin film of $MgB_2$ is grown on an R-plane sapphire substrate as described elsewhere.[14] The superconducting transition temperature ($T_c$) onset is 39 K, and the 10 – 90% transition width is 0.7 K, comparable to high quality bulk samples. A high density sintered polycrystalline pellet of $MgB_2$, also with $T_c$ = 39 K, is synthesized at high pressure and characterized as described elsewhere.[6,15] The polycrystalline pellet contains $MgB_4$ as a minor phase detected in transmission electron microscopy,[15] but is 99% $MgB_2$ as determined by x-ray diffraction.[6] Optical quality single crystals of MgO and $MgF_2$ with (100) orientation are also studied for comparison. The sample surfaces are cleaned in a dry box with an inert ultrahigh purity $N_2$ atmosphere which encloses the load lock area of the XPS spectrometer. An etchant consisting of 1% HCl in absolute ethanol or 0.5% $Br_2$ in absolute ethanol is used for the $MgB_2$ and MgO, and 10% HF in absolute ethanol is used for the $MgF_2$. The samples are immersed in the etchant for 15-60 seconds, rinsed in ethanol, blown dry with nitrogen, and loaded into the XPS spectrometer with no atmospheric exposure. The etchant removes surface oxides, hydroxides, and carbonates by the formation of reaction products, in this case $MgCl_2$ and $BCl_3$ or $MgBr_2$ and $BBr_3$, which are soluble in nonaqueous solvents such as alcohols. Other than the small amount of water originating from the concentrated HCl, the etchants used for $MgB_2$ are nonaqueous. Nonaqueous etching has previously been shown to yield high quality surfaces of high temperature superconductors,[16,17] which also contain reactive alkaline earth elements.

The XPS spectrometer is a Surface Science Instruments SSX-501 utilizing monochromatic Al $K_\alpha$ x-rays (1486.6 eV). Spectra are measured at ambient temperature with photoemission 55° from the surface normal for the $MgB_2$ pellet and normal to the surface for the film. Data from both samples are qualitatively similar except as otherwise noted. The data presented in the figures are from the film. Charging of the insulating MgO and $MgF_2$ samples is compensated with an electron flood gun and the spectra are referenced to the adventitious C 1s signal at 284.6 eV. The spectra are measured with an x-ray spot size of 150 or 300 μm and the pass energy of the electron energy analyzer is set to 25 eV. The energy scale is calibrated using sputter-cleaned Au and Cu with the Au $4f_{7/2}$ binding energy set to 83.95 ± 0.05 eV (0.80 eV full width at half-maximum (FWHM)) and the Cu $2p_{3/2}$ binding energy set to 932.45 ± 0.05 eV (0.97 eV FWHM).

## 3. Results and Discussion

The XPS core level binding energies and Mg $KL_{23}L_{23}$ Auger kinetic energies measured in this work are summarized in Table 1. The as-grown $MgB_2$ surface exhibits significant levels of C and O in addition to Mg and B signals. The Mg 2p core levels and Mg $KL_{23}L_{23}$ Auger transitions before and after etching are shown in Figures 1 and 2, respectively. Equivalent results are obtained for the HCl and $Br_2$ etchants, but the $Br_2$ etchant leaves a small residue with a Br 3p signal which overlaps with the B 1s signal, while the HCl etchant leaves less residue and any Cl signals do not overlap with signals of interest. Measurements from etched surfaces presented here are therefore from HCl-etched surfaces. Two Mg 2p signals are evident in Figure 1, and two Mg $KL_{23}L_{23}$ Auger signals are also observed in Figure 2 more clearly due to the greater energy separation. The reduced intensities of the high binding energy (low kinetic energy) peaks after etching correlates with the elimination of a carbonate C 1s signal (288.5 eV) and reduction of the O 1s signal (532.4 eV). The lower binding energy Mg 2p (49.35 eV, FWHM 0.67 eV) and higher kinetic energy Mg $KL_{23}L_{23}$ (1185.0 eV) signals are therefore assigned to the $MgB_2$ phase, and are slightly lower in energy compared to those observed for Mg metal (Mg 2p at 49.8 eV and Mg $KL_{23}L_{23}$ at 1185.5 eV).[18] The O 1s binding energy is significantly higher than that measured from the MgO (100) surface, and most likely originates from $MgCO_3$ and $Mg(OH)_2$.

The B 1s signals measured before and after chemical etching do not differ qualitatively, as shown in Figure 3. The B 1s binding energy of the major signal (186.55 eV, FWHM 0.85 eV) is somewhat lower than measurements from transition metal diborides, which have B 1s binding energies in the range 187.2 – 188.5 eV.[19,20] In addition to the major signal, a minor peak near 192 eV is present, consistent with $B_2O_3$. The B 1s signal measured from the polycrystalline pellet differs from the data in Figure 3 in having an additional signal at 187.9 eV.

The as-grown surface of the $MgB_2$ polycrystalline pellet, in addition to significant levels of C and O, exhibits a Mg:B ratio of 1:1.25. Attempts to obtain clean surfaces by mechanical abrasion were unsuccessful, yielding spectra which did not differ significantly from the as-grown surface. This may indicate that intragrain fracturing is rare, i.e. abrasion primarily exposes fresh grain boundary surfaces. The greater reactivity of Mg relative to B results in the observed Mg-rich oxidized surface layer, which is also observed on the film. Similar observations are made for surfaces of alkaline earth-containing high temperature superconductors.[17]

Increasing the chemical etch time to 60s did not significantly change the measured spectra compared to a 15s etch time. The Mg:B ratio of the polycrystalline pellet surface after etching is 1:3, indicating that the subsurface region is depleted of Mg. This Mg-deficient layer is not observed for the film. The Mg:B ratio does not change with increasing etch time, suggesting that the etchant removes only the oxidized surface layer and does not appear to leach Mg from the bulk or to etch unoxidized B.  The presence of a reacted Mg surface layer on a Mg-deficient layer for the polycrystalline pellet may explain the variation in results obtained with surface-sensitive STS measurements.[9-13]  A Mg-deficient surface layer would have lower electron density in the B planes, consistent with the higher binding energy B 1s signal measured from the polycrystalline pellet.

The Mg 2p core level of chemically-etched $MgB_2$ is compared to measurements from MgO and $MgF_2$ in Figure 4.  It is interesting to note that the Mg 2p signals of both $MgB_2$ and MgO occur at lower binding energies than that of Mg metal, despite the +2 charge of the Mg ion in both compounds.  Other alkaline earth compounds also exhibit this unusual negative chemical shift,[21] which has been attributed to initial-state electrostatic effects primarily due to the large value of the Madelung energy relative to the ionization energy.[21-24]  The Mg $KL_{23}L_{23}$ Auger signals are compared in Figure 5.  From the Auger parameter (sum of the Mg core level binding energy and Auger transition kinetic energy) differences, the final state relaxation energy relative to $MgF_2$ can be estimated to be 0.9 eV larger in MgO and 2.6 eV larger in $MgB_2$.  The larger value of the relaxation energy for MgO is consistent with the larger polarizability of $O^{2-}$ ions relative to $F^-$ ions, resulting in greater polarization screening of core holes.  The large value of the relaxation energy for $MgB_2$ is nearly identical to that of Mg, and is consistent with metallic screening of core holes.  This observation is consistent with calculations showing that electrons in B $2p_z$ states are delocalized (i.e. metallic conductivity between B planes) and Mg-Mg bonding is metallic.[4,5,25,26]

The measured $MgB_2$ valence band is shown in Figure 6.  A single well-defined peak near 6 eV is observed, with less well-defined structure at higher and lower binding energies.

There is intensity observed at the Fermi level and the suggestion of a Fermi edge, but within the resolution and signal-to-noise ratio of the measurement a clear Fermi edge is not unambiguously apparent. The total density of states (DOS) determined from band structure calculations[2] is shown below the measured spectrum. A good correspondence with the measured spectrum is obtained if the DOS is shifted to higher binding energy by 4 eV, as shown in Figure 6. Similar shifts of ~2 eV are required for high temperature superconductors, and have been attributed to electron correlation effects.

**4. Summary and Conclusions**

A c-axis oriented film and a high density sintered pellet of $MgB_2$ have been studied by XPS, and the results have been compared to measurements from MgO and $MgF_2$ single crystals. The as-grown surface is found to be Mg-rich and oxidized, resulting from the greater reactivity of Mg relative to B. This oxidized layer is effectively removed by nonaqueous etchants of 1% HCl in absolute ethanol or 0.5% $Br_2$ in absolute ethanol. The subsurface region of the sintered pellet is found to be Mg-deficient. This reacted surface layer and nonstoichiometric subsurface region may explain the varied results obtained in surface-sensitive scanning tunneling spectroscopy (STS) measurements.[9-13] Indeed, STS measurements are significantly improved by the chemical etching described in this work, exhibiting improved spatial homogeneity and a nearly vanishing DOS near $E_F$.[13] The Mg core level and Auger signals of $MgB_2$ are similar to measurements from metallic Mg, the B 1s signal is similar to measurements from transition metal diborides, and the measured valence band is consistent with the calculated density of states.

**Acknowledgments**

Part of the work described in this paper was performed by the Center for Space Microelectronics Technology, Jet Propulsion Laboratory, California Institute of Technology, and was sponsored by the National Aeronautics and Space Administration. Part of this work was performed at Pohang University of Science and Technology, and supported by the Ministry of Science and Technology of Korea through the Creative Research Initiative Program.

**Table 1:** Summary of XPS core level binding energies and Mg $KL_{23}L_{23}$ Auger kinetic energies measured in this work.

| Material | Mg 2p | Mg KLL | Anion 1s |
|---|---|---|---|
| $MgB_2$ | 49.35 | 1185.0 | 186.55 (B 1s) |
| MgO | 48.9 | 1182.0 | 529.3 (O 1s) |
| $MgF_2$ | 51.0 | 1178.1 | 685.5 (F 1s) |

# Figure Captions

1. Mg 2p spectra measured from a c-axis oriented $MgB_2$ film (a) with no surface treatment, and (b) after nonaqueous chemical etching.

2. Mg $KL_{23}L_{23}$ Auger transitions measured from a c-axis oriented $MgB_2$ film (a) with no surface treatment, and (b) after nonaqueous chemical etching.

3. B 1s spectra measured from a c-axis oriented $MgB_2$ film (a) with no surface treatment, and (b) after nonaqueous chemical etching.

4. Comparison of the Mg 2p spectra measured from a c-axis oriented $MgB_2$ film and (100) single crystal surfaces of MgO and $MgF_2$.

5. Comparison of the Mg $KL_{23}L_{23}$ Auger transitions measured from a c-axis oriented $MgB_2$ film and (100) single crystal surfaces of MgO and $MgF_2$.

6. Comparison of the measured $MgB_2$ valence band (top curve) with the total density of states (DOS) calculated in Ref. 2 (bold curve). A shift of the DOS to higher binding energy by 4 eV is necessary to approximate the measured spectrum.

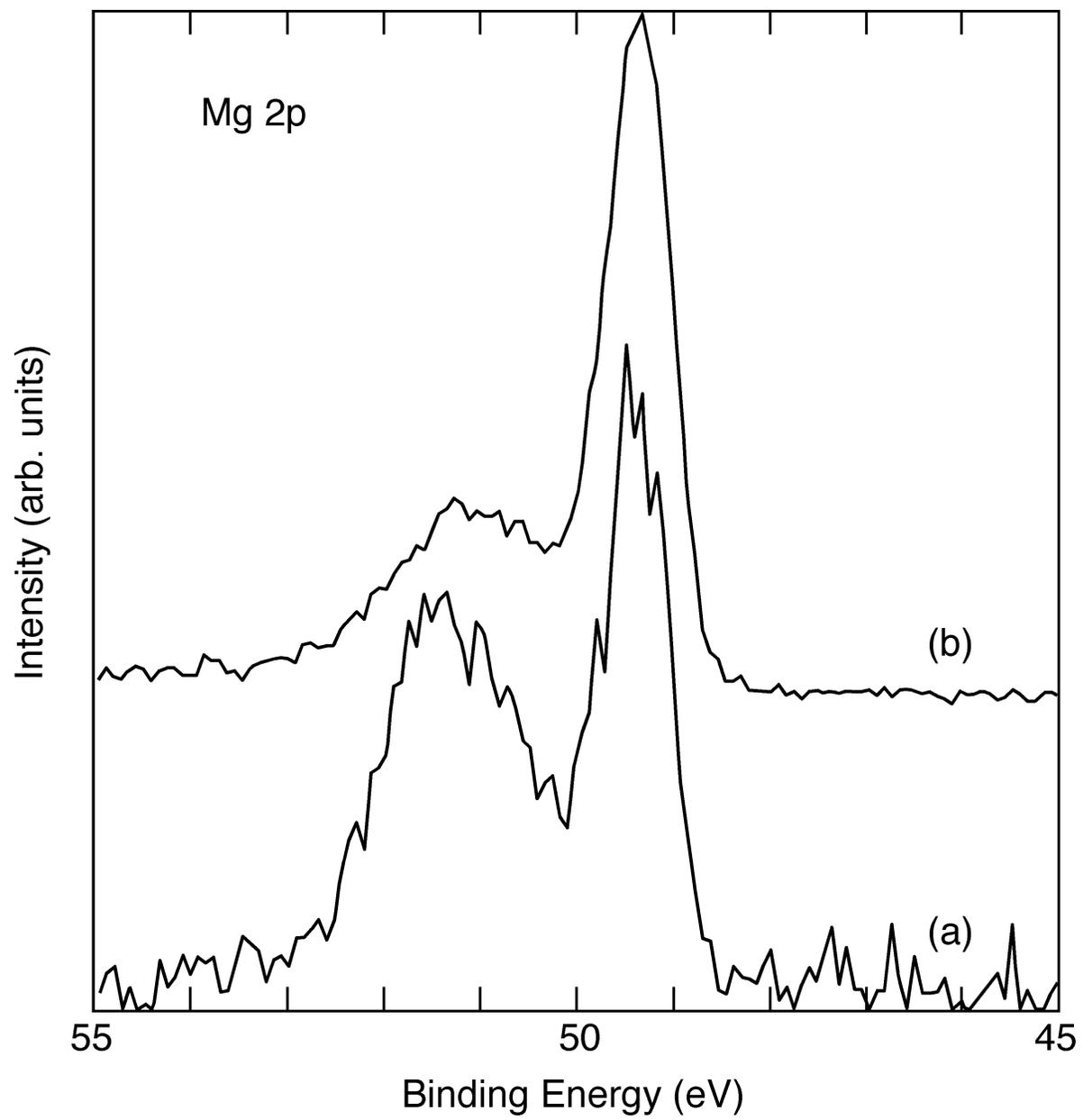

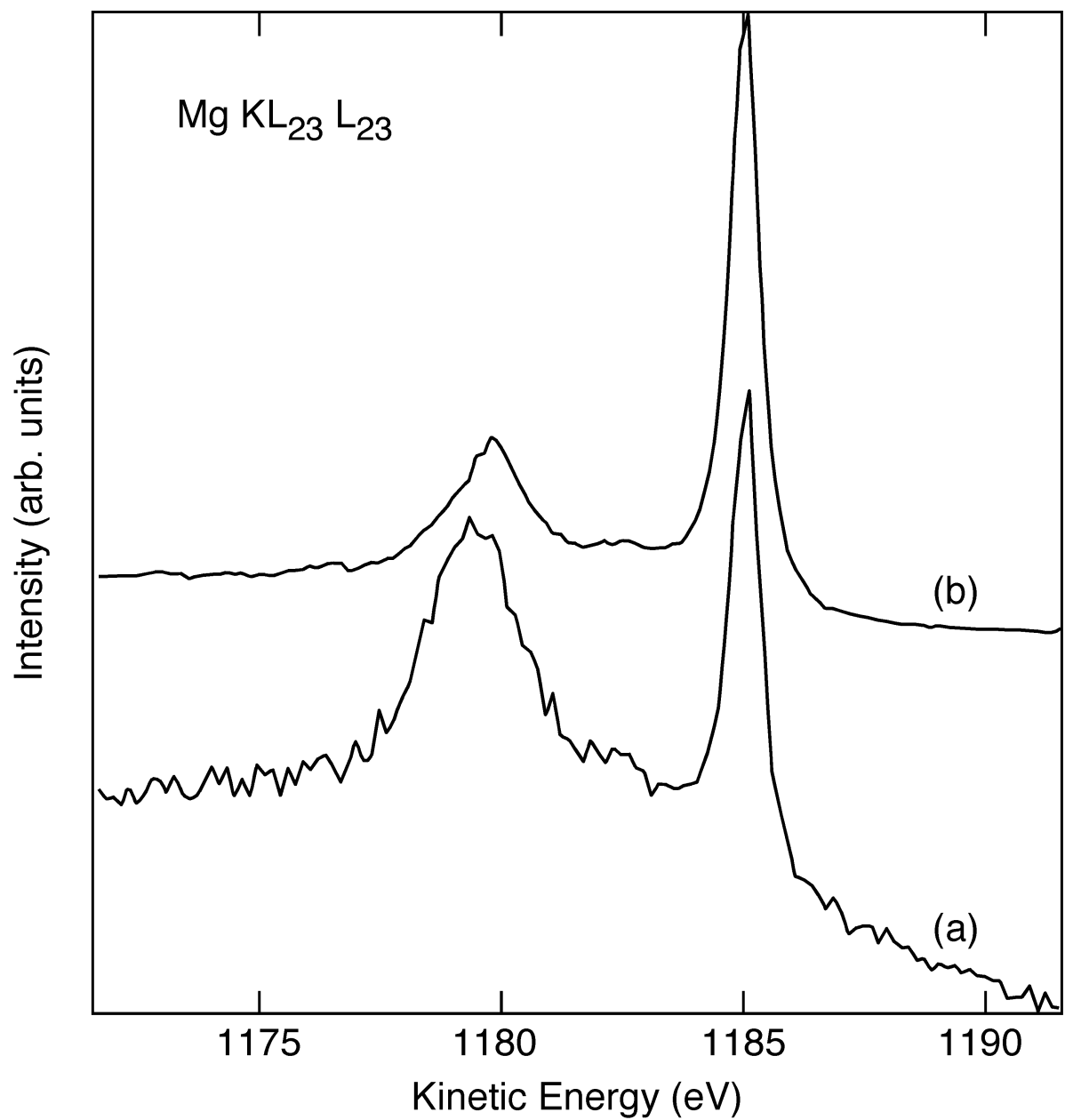

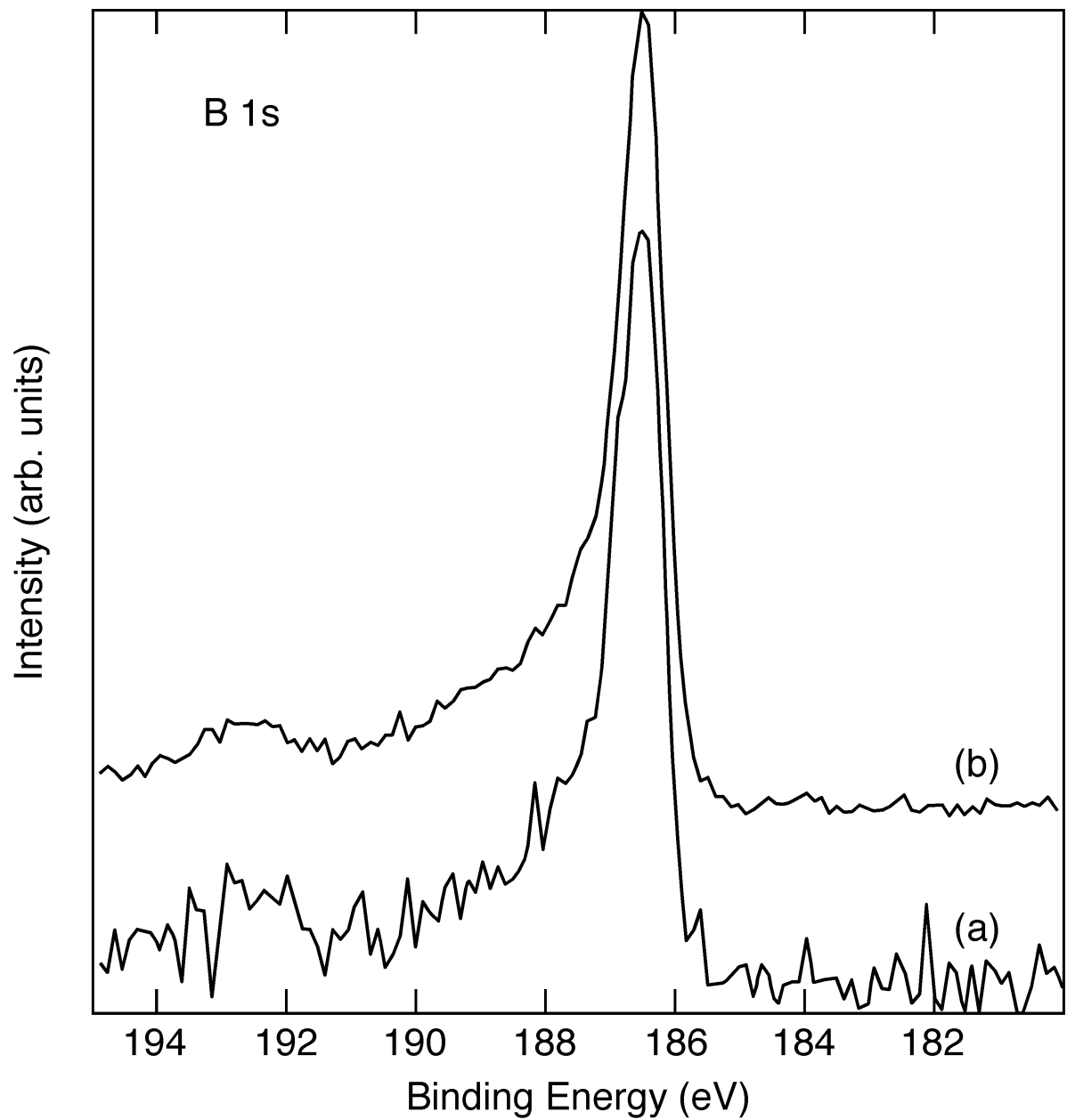

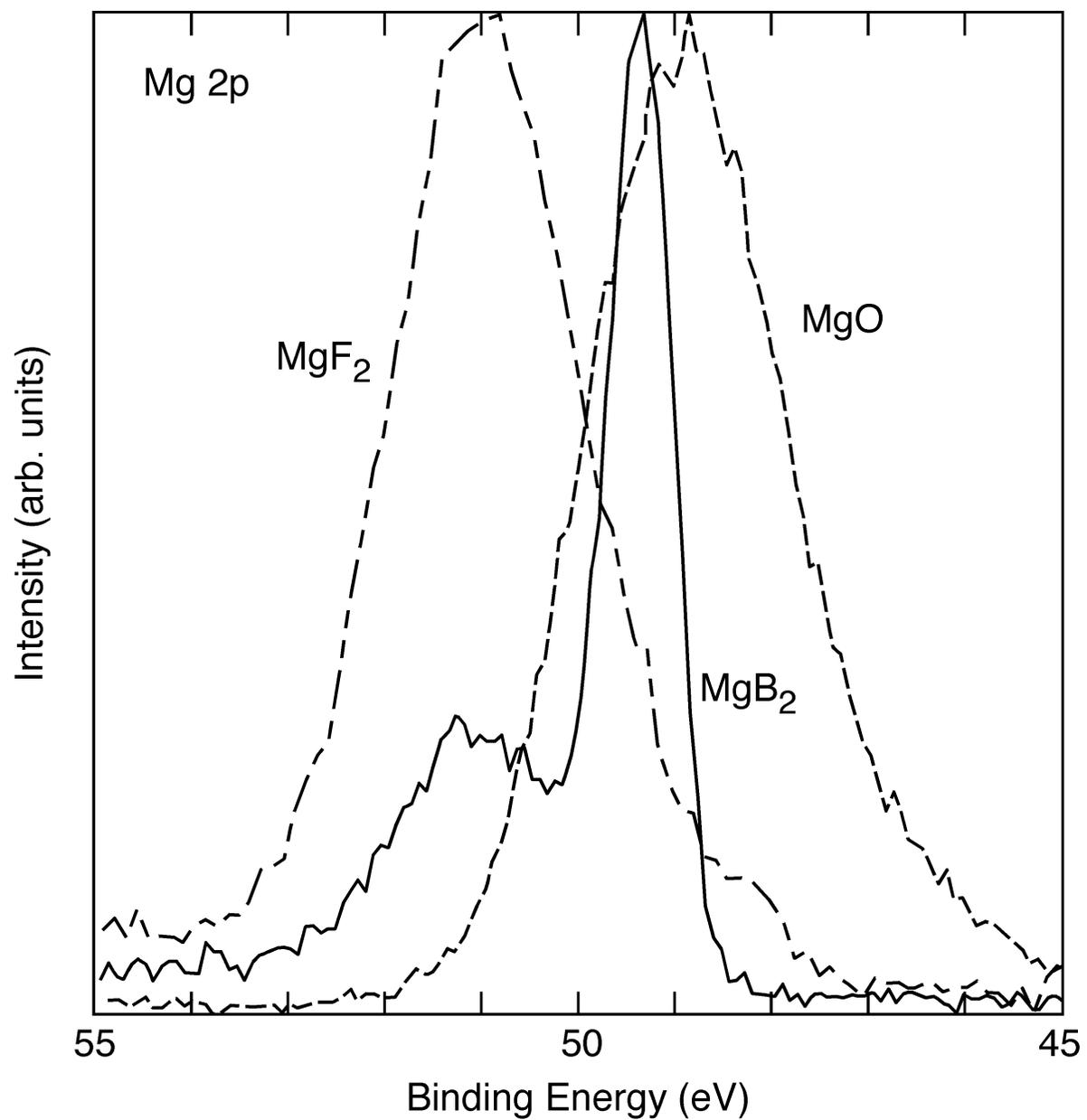

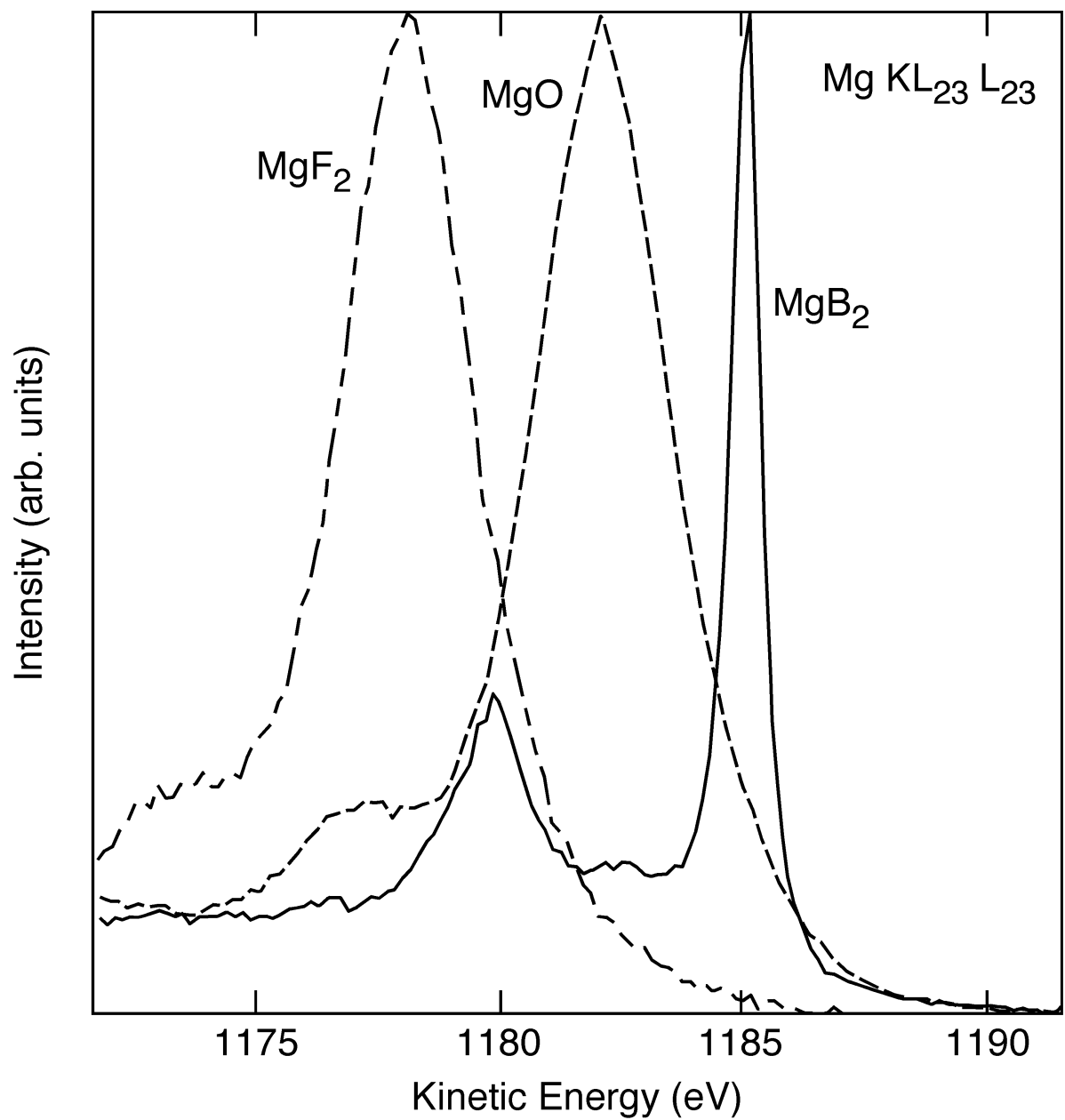

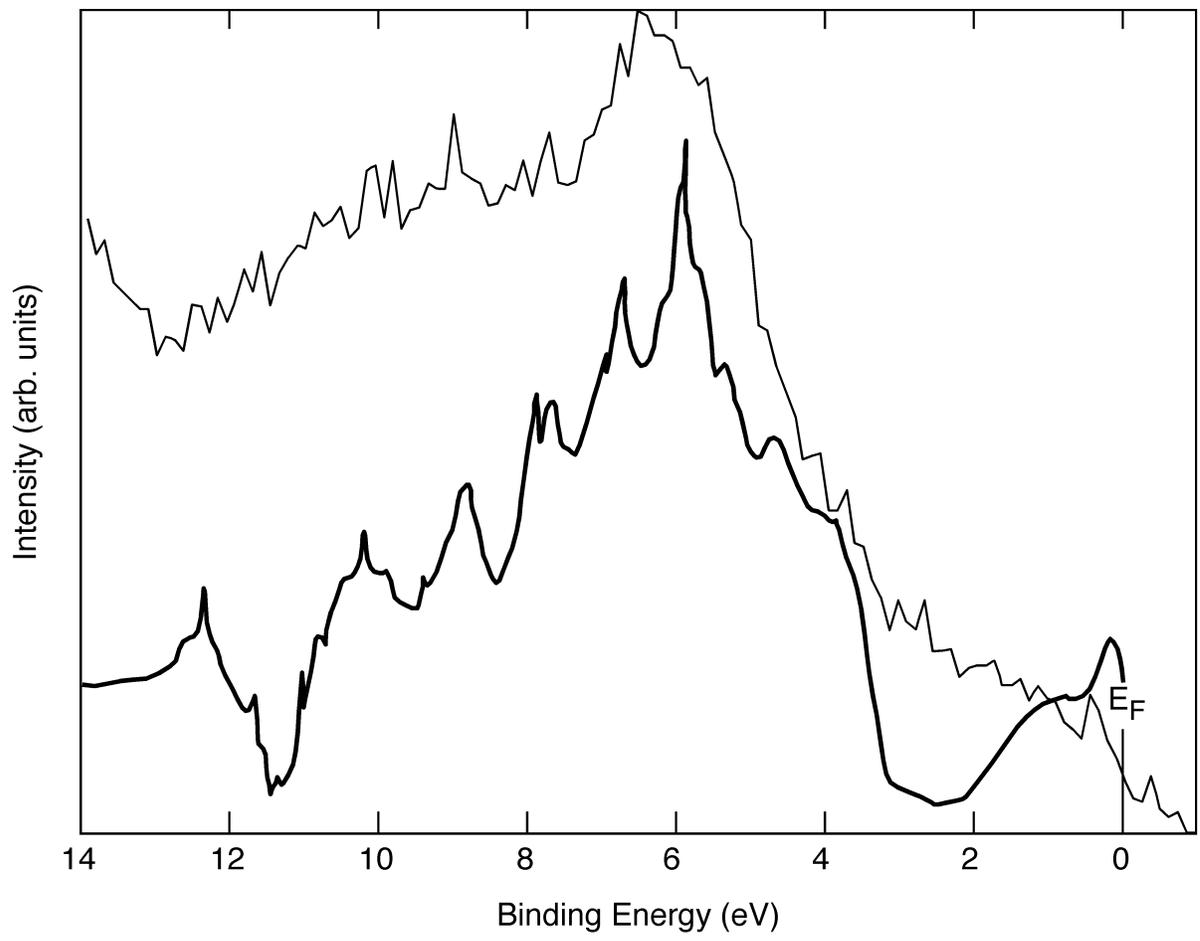